\documentclass[a4paper]{jpconf}
\usepackage{graphicx}
\usepackage{citesort}
\begin{document}
\title{Unveiling new systematics in the self-assembly of atomic chains on Si(111)}

\author{C. Battaglia, H. Cercellier, C. Monney, L. Despont, M.G. Garnier,\\ P. Aebi}
\address{Institut de Physique, Universit\'e de Neuch\^atel, 2000 Neuch\^atel, Switzerland}%
\ead{corsin.battaglia@unine.ch}

\begin{abstract}
Self-assembled arrays of atomic chains on Si(111) represent a
fascinating family of nanostructures with quasi-one-dimensional
electronic properties. These surface reconstructions are stabilized
by a variety of adsorbates ranging from alkali and alkaline earth
metals to noble and rare earth metals. Combining the complementary
strength of dynamical low-energy electron diffraction, scanning
tunneling microscopy and angle-resolved photoemission spectroscopy,
we recently showed that besides monovalent and divalent adsorbates,
trivalent adsorbates are also able to stabilize silicon honeycomb
chains. Consequently silicon honeycomb chains emerge as a most
stable, universal building block shared by many atomic chain
structures. We here present the systematics behind the self-assembly
mechanism of these chain systems and relate the valence state of the
adsorbate to the accessible symmetries of the chains.
\end{abstract}

\section{Introduction}

Stimulated by their great potential for next generation electronic,
magnetic, optical and chemical devices, research on ordered
nanostructures has attracted tremendous interest. Their fabrication
remains a challenging task. Spontaneous self-assembly represents a
promising bottom-up approach, which allows to fabricate
nanostructures in a massively parallel fashion. Understanding the
mechanisms underlying spontaneous nanostructure formation is crucial
for establishing control over their dimensions, spatial distribution
and uniformity.\\
Here we address the self-assembly of macroscopic arrays of atomic
chains on silicon surfaces which have been the focus of intense
research because of their fascinating quasi-one-dimensional
 electronic properties \cite{Crain03,Ahn05,Guo05,Snijders06}.
A large variety of adsorbates ranging from alkali and alkaline earth
metals to noble and rare earth metals is known to induce a
reconstruction of the Si(111) surface into atomic chains. Most chain
structures are based on a common structural backbone formed by
silicon atoms. The role of the adsorbate is to stabilize these
silicon chains by donating the correct number of electrons. We
recently showed that besides monovalent and divalent adsorbates,
trivalent adsorbates are also able to stabilize Si atomic chains
\cite{Battaglia07}. In this paper we first review the structure for
the monovalent and divalent adsorbates and then introduce the
extension to the trivalent adsorbates. This new piece of information
allows us to understand new aspects of the systematics behind the
self-assembly process, from which we can draw conclusions about the
accessible symmetries of the chains.

\begin{figure}[h]
\includegraphics[width=20pc]{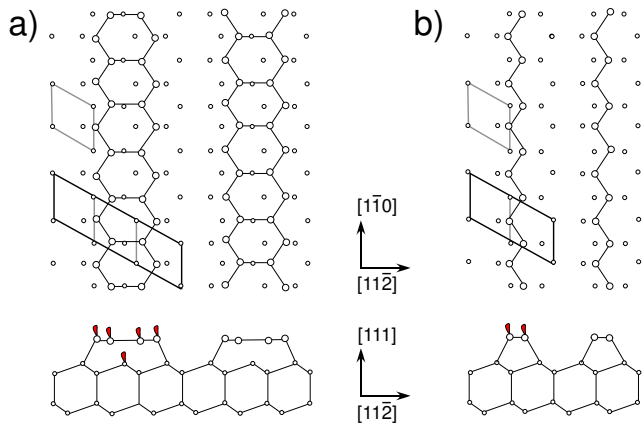}\hspace{2pc}%
\begin{minipage}[b]{14pc}\caption{\label{Fig1}{\small Top (upper part) and side (bottom part) view of the two prototypical silicon chains separated by the channels which accommodate the adsorbate atoms (not drawn): a) honeycomb chain with (3$\times$1) unit cell and b) Seiwatz chain with (2$\times$1) unit cell. The (1$\times$1) unit cell of the (111) oriented substrate is shown in grey. The dangling bonds of under-coordinated silicon atoms are schematically indicated in the side view by the red symbols. The crystallographic directions of the substrate are also drawn.}}
\end{minipage}
\end{figure}

\section{Monovalent adsorbates}

Monovalent adsorbates such as alkali metals (Li, Na, K, Rb, Cs)
together with Ag are known to induce a chain reconstruction on
Si(111) exhibiting a (3$\times$1) superstructure. The widely
accepted structural model, the so-called honeycomb chain-channel
(HCC) model \cite{Lottermoser98,Collazo-Davila98,Erwin98}, is shown
in Fig. \ref{Fig1}a). The (3$\times$1) unit cell (black) and its
relation to the (1$\times$1) surface unit cell (grey) of the (111)
oriented substrate is shown as well. The HCC model consists of Si
honeycomb chains running along the $[1\overline{1}0]$ direction
separated by empty channels. The structure contains five threefold
coordinated, thus under-coordinated, silicon atoms per (3$\times$1)
unit cell. Each of the five associated silicon dangling bonds,
marked schematically by the red symbols in Fig. \ref{Fig1}a), gives
rise to a surface state. A total of five electrons, one from each
dangling bond, is available to fill these states. Since each surface
state is able to accommodate a maximum of exactly two electrons, we
get two completely filled surface states below the Fermi energy, one
half-filled metallic surface state crossing the Fermi energy and two
empty states. However, in order to stabilize the HCC structure one
more electron is required. This additional electron, which is
provided by the adsorbate atom, allows an enormous energy gain by
completely filling the metallic state, rendering it insulating by
bringing it below the Fermi energy \cite{Erwin98}. The adsorbate
atoms, represented by blue disks in Fig. 2a), are occupying the
sites inside the channels. Since one electron per (3$\times$1) unit
cell is required, there is one monovalent adsorbate per (3$\times$1)
unit cell and thus the adsorbate coverage is 1/3 monolayer (ML).

\begin{figure}[h]
\includegraphics[width=38pc]{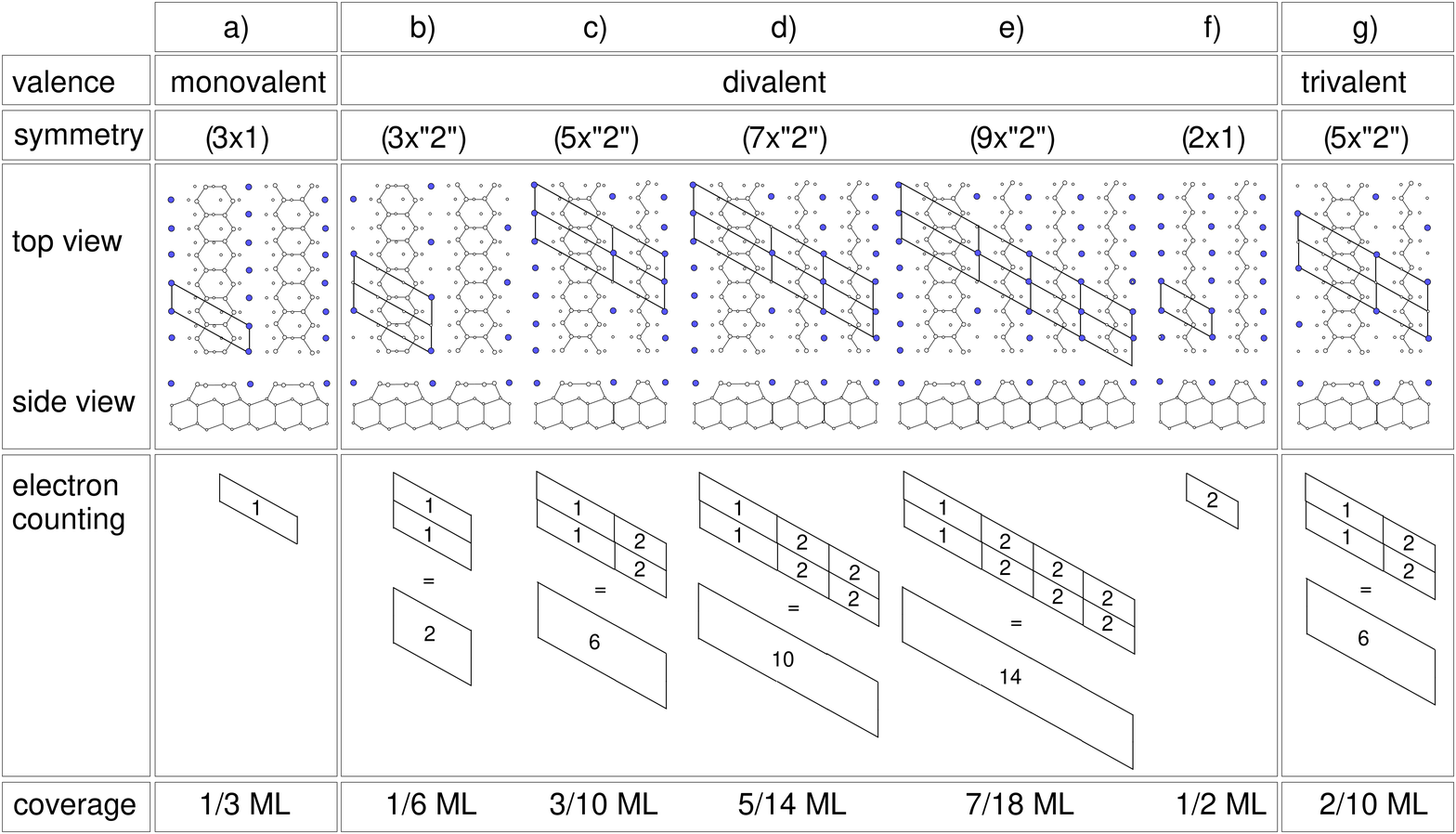}\hspace{2pc}%
\caption{\label{Fig2}{\small Systematics of adsorbate induced
silicon chains. Adsorbate atoms are drawn in blue. Valence of the
adsorbate, symmetry of the unit cell, top and side view of the
atomic structure are shown. The line labeled electron counting,
shows how the final unit cell can be decomposed into the basic
building blocks. The number inside each unit cell gives the number
of electrons required to stabilize each building block. The last
line gives the adsorbate coverage for each structure.}}
\end{figure}

\section{Divalent adsorbates}

Silicon honeycomb chains may also be stabilized by divalent
adsorbates such as the alkaline earth metals (Mg, Ca, Sr, Ba) and
rare earth metals (Sm, Eu, Yb) \cite{Sakamoto05b}. Although these
rare earth metals commonly occur in a +3 valence state, a +2
configuration is occasionally preferred as in this case. Since a
divalent adsorbate provides twice as many electrons compared to a
monovalent adsorbate, only 1/6 ML, i.e. half the monovalent
adsorbate coverage is required \cite{Lee01}. Consequently only every
second site in the channel is occupied leading to a doubling of the
unit cell size along the chain direction (see Fig. \ref{Fig2}b).
Thus for divalent adsorbates the unit cell has a (3$\times$2)
symmetry. In LEED patterns however, instead of sharp $\times$2
spots, only faint half-order streaks parallel to the $\times3$ spots
are observed. These are caused by the occurrence of two consecutive
empty sites in the channel. Such a defect leads to a registry shift
of the adsorbate sequence in one channel with respect to adsorbates
in the adjacent channel by one period along the chain direction.
Since adsorbates in adjacent channel are only very weakly coupled,
the two configurations are energetically almost degenerate. The
local mixing of these two arrangements with poor long range order is
responsible for the $\times$2 streaks seen in LEED patterns. To
indicate the presence of half-order streaks one uses the notation
(3$\times$"2") where the $\times$"2" stands for a $\times$2
periodicity with missing coherence between adjacent chains \cite{Sakamoto02}. \\
Besides the prototypical silicon honeycomb chains, divalent
adsorbates induce a second type of silicon chain structure at higher
coverage, the so-called Seiwatz chains \cite{Baski01,Sekiguchi01}
characterized by a (2$\times$1) unit cell as shown in Fig.
\ref{Fig1}b). Seiwatz chains consist of zig-zag chains of silicon
atoms separated by empty channels. Seiwatz chains require two
electrons per (2$\times$1) unit cell, since they are stabilized by
1/2 ML of divalent adsorbates. Consequently every available site in
the channel must be occupied by a divalent adsorbate as shown in
Fig. \ref{Fig2}f). Monovalent adsorbates instead are not able to
stabilize Seiwatz chains, since they are not able to
donate the required amount of electrons.\\
At intermediate coverages, divalent adsorbates stabilize chain
structures with (5$\times$"2"), (7$\times$"2") and even
(9$\times$"2") depending on the adsorbate. These are considered to
be composed of an appropriate combination of honeycomb chains and
Seiwatz chains. For the (5$\times$"2") symmetry for instance shown
in Fig. \ref{Fig2}c), a honeycomb chain alternates with a Seiwatz
chain. The (5$\times$"2") unit cell can be thought of as being
composed of two honeycomb unit cells with (3$\times$1) symmetry plus
two Seiwatz unit cells with (2$\times$1) symmetry as schematized in
Fig. \ref{Fig2}c). The adsorbates have to donate one electron for
each of the two honeycomb units plus two electrons for each of the
two Seiwatz units. Thus the adsorbates have to provide a total of
six electrons per (5$\times$"2") unit cell. This condition is easily
satisfied by placing three divalent adsorbates per (5$\times$"2") in
between the Si chains resulting in a coverage of 3/10 ML. For the
(7$\times$"2") and the (9$\times$"2")
 symmetry shown in Fig. \ref{Fig2}d) and \ref{Fig2}e)  a honeycomb
chain alternates with two respectively three successive Seiwatz
chains, requiring the adsorbates to provide 10 respectively 14
electrons, resulting in an adsorbate coverage of 5/14 respectively
7/18 ML. Note that these intermediate symmetries are not accessible
to monovalent adsorbates, since it requires the stabilization of
Seiwatz chains.

\section{Trivalent adsorbates}

For trivalent adsorbates such as Gd \cite{Kirakosian02}, but also Dy
\cite{Engelhardt06}, Er \cite{Wetzel97}, Ho \cite{Himpsel04} chain
reconstructions with (5$\times$"2") symmetry have been observed. To
construct the (5$\times$"2") unit cell, we combine once again a
honeycomb chain with a Seiwatz chain (see Fig. \ref{Fig2}g). The
required six electrons are easily obtained by placing two trivalent
adsorbates per unit cell in the channels. Note that for the divalent
adsorbates, we had to take three adsorbates to satisfy the doping
criterion. The resulting coverage for the trivalent adsorbates is
2/10 ML, a value which is consistent with experiment
\cite{Kirakosian02}. It is important to emphasize that besides the
(5$\times$"2") symmetry no other symmetry has been observed for the
trivalent adsorbates. Our model gives an easy explanation: the
(7$\times$"2") and the (9$\times$"2") symmetry require 10
respectively 14 electrons to be stabilized - a condition which can
not be satisfied by a trivalent adsorbate. Thus electron counting
provides an intuitive picture for the occurrence of the various
allowed symmetries.

\section{Conclusion}

Self-assembly of atomic chains on silicon surfaces is driven by the
elimination of dangling bonds. Most chain structures share a common
building block formed by silicon atomic chains. The fact that only
silicon atoms participate in the formation of the chains allows a
variety of adsorbates to adopt these one-dimensional nanostructures.
Using a simple electron counting model, we are able to relate the
valence state of the adsorbate to the accessible symmetries of the
chains.




 \ack
Stimulating discussions with Christian Koitzsch and Pascal Ruffieux
are gratefully acknowledged. Skillfull technical assistance was
provided by our workshop and electric engineering team. This work
was supported by the Fonds National Suisse pour la Recherche
Scientifique through Div. II and MaNEP.

\section*{References}
 \bibliography{GdSi111}
\enlargethispage{1cm}




\end{document}